# Generalized Kasami Sequences: The Large Set

Xiangyong Zeng, Qingchong Liu, *Member*, Lei Hu

## Abstract

In this paper new binary sequence families $\mathcal{F}^k$ of period $2^n - 1$ are constructed for even $n$ and any $k$ with $\gcd(k, n) = 2$ if $n/2$ is odd or $\gcd(k, n) = 1$ if $n/2$ is even. The distribution of their correlation values is completely determined. These families have maximum correlation $2^{n/2+1} + 1$ and family size $2^{3n/2} + 2^{n/2}$ for odd $n/2$ or $2^{3n/2} + 2^{n/2} - 1$ for even $n/2$. The construction of the large set of Kasami sequences which is exactly the $\mathcal{F}^k$ with $k = n/2 + 1$ is generalized.

## Index Terms

Periodic correlation, $m$-sequence, low correlation, Kasami sequence, quadratic form

X. Zeng and Q. Liu's work is supported in part by NSF under Grant CNS-0435341. L. Hu's work is supported by the National Science Foundation of China (NSFC) under Grants No. 60373041 and No. 60573053.

X. Zeng and Q. Liu are with Department of Electrical and System Engineering, Oakland University, Rochester, MI 48309, USA. Email: {zeng2, qliu}@oakland.edu.

L. Hu is with the Graduate School of Chinese Academy of Sciences, 19A Yuquan Road, Beijing 100049, P. R. China. Email: hu@is.ac.cn.

## I. INTRODUCTION

Binary sequences are traditionally employed by CDMA systems, spread spectrum systems and broadband satellite communications [1]. Families of sequences for such applications are desired to have low autocorrelation, low cross-correlation, and large family size [2].

The maximal length binary sequences ($m$-sequences) have a simple representation based on the trace function $tr_1^n(\cdot)$. Since $m$-sequences have ideal autocorrelation properties, it is natural to study the cross-correlation function between an $m$-sequence and its decimations. Many families of low correlation sequences have been constructed by using $m$-sequences and their decimations [2], [3]. For instance, the well-known Gold sequence family [4] constructed from a pair of $m$-sequences

$$\{tr_1^n(x)\} \text{ and } \{tr_1^n(x^{2^k+1})\} \tag{1}$$

for odd $n$ and integer $k$ with $\gcd(n,k) = 1$. The small set of Kasami sequences [5], [6] constructed from

$$\{tr_1^n(x)\} \text{ and } \{tr_1^n(x^{2^{n/2}+1})\} \tag{2}$$

for even $n$. These families have the maximum correlation $2^{(n+1)/2} + 1$ or $2^{n/2} + 1$.

The Gold-like sequences constructed by combining sequences

$$\{tr_1^n(x)\} \text{ and } \Big\{ \sum_{k=1}^{(n-1)/2} tr_1^n(x^{2^k+1}) \Big\} \tag{3}$$

by Boztas and Kumar for odd $n$ [7] are another example of good sequence sets. Its analogy for even $n$ was introduced by Udaya [8]. These constructions were further generalized by Kim and No [9]. They can achieve a larger linear span than Gold sequences, but have the same family size as the latter. The construction was extended to the nonbinary case in [10].

Combining an $m$-sequence and its decimations is an efficient method to construct sequence families with larger family size and large linear span. The large set of Kasami sequences derived from the sequences

$$tr_1^n(x), \ tr_1^{n/2}(x^{2^{n/2}+1}) \text{ and } tr_1^n(x^{2^{n/2+1}+1}) \tag{4}$$

for even $n$ is such a family [5], [11]. It has the maximum correlation of $2^{n/2+1} + 1$ and the family size of $2^{3n/2} + 2^{n/2}$ or $2^{3n/2} + 2^{n/2} - 1$. The modified Gold codes [12] can achieve a much larger family size. However, it is difficult to determine the correlation distribution.

In this paper, we generalize the construction of the large set of Kasami sequences. We assume $n$ is even and take $k$ to satisfy $\gcd(k, n) = 2$ if $n/2$ is odd and $\gcd(k, n) = 1$ if $n/2$ is even. We then combine three $m$-sequences

$$tr_1^n(x), \ tr_1^n(x^{2^k+1}) \ \text{and} \ tr_1^{n/2}(x^{2^{n/2}+1}) \tag{5}$$

to get a family of binary sequences with same family size and correlation distribution as the large set of Kasami sequences. The proposed families include the small set of Kasami sequences as their subfamily. When $k = n/2 + 1$, this family is the large set of Kasami sequences.

This paper focuses on the correlation distribution of the proposed families. The work is closely related to determining the number of solutions to the equation

$$\varepsilon x^{2^l+1} + \upsilon x + \theta = 0 \tag{6}$$

and the number of elements $\theta \in F_{2^n}$ such that the equation

$$\theta^{2^{n-k}} z^{2^{n-k}} + \theta z^{2^k} + z^{2^{n/2}} = 0 \tag{7}$$

has nonzero solutions, where $\gcd(l, n) = 1$, $\varepsilon$, $\theta$, and $\upsilon$ are elements in $F_{2^n}$, and $\varepsilon$ and $\theta$ are not zero. To achieve this goal, a linear code and its dual are introduced.

The remainder of this paper is organized as follows. Section II gives some definitions and preliminaries. Sections III studies the equations mentioned above. Section IV characterizes the Fourier transform for a class of quadratic functions. In Section V we propose the generalized sequence sets and determine their correlation distribution. Section VI concludes the study.

## II. PRELIMINARIES

Let $\mathcal{F}$ be the family of $M$ binary sequences of period $2^n - 1$ given by

$$\mathcal{F} = \{\{s_i(t)\}_{t=0}^{2^n-2} \mid 0 \le i \le M-1\}. \tag{8}$$

The *periodic correlation function* of the sequences $\{s_i(t)\}$ and $\{s_j(t)\}$ in $\mathcal{F}$ is

$$R_{i,j}(\tau) = \sum_{t=0}^{2^n-2} (-1)^{s_i(t)-s_j(t+\tau)} \tag{9}$$

where $0 \le i, j \le M-1$, and $0 \le \tau \le 2^n - 2$. The *maximum magnitude* $R_{\max}$ of the correlation values is

$$R_{\max} = \max |R_{i,j}(\tau)|$$

where $0 \leq i, j \leq M - 1$, $0 \leq \tau \leq 2^n - 2$, and the cases of in-phase autocorrelation ($i = j$ and $\tau = 0$) are excluded.

Let $A_i$ denote the number of codewords of weight $i$ in a linear binary $[m, k]$ code $\mathcal{C}$. The polynomial

$$\sum_{i=0}^{m} A_i x^{m-i} y^i,$$

denoted by $W_{\mathcal{C}}(x, y)$, is called the *weight enumerator* of $\mathcal{C}$. It is known in the theory of coding [13] that the dual code $\mathcal{C}^{\perp}$ of $\mathcal{C}$ has the weight enumerator as

$$W_{\mathcal{C}^{\perp}}(x, y) = 2^{-k} W_{\mathcal{C}}(x + y, x - y). \tag{10}$$

Let $F_{2^n}$ be the finite field with $2^n$ elements, and $n = ml$ for integers $m$ and $l$. The *trace function* $tr_m^n(\cdot)$ from $F_{2^n}$ to $F_{2^m}$ is defined by

$$tr_m^n(x) = \sum_{i=0}^{l-1} x^{2^{im}}, x \in F_{2^n}.$$

The trace function has the following properties [14]:

i) $tr_m^n(ax + by) = a \, tr_m^n(x) + b \, tr_m^n(y)$ for any $a, b \in F_{2^m}$ and any $x, y \in F_{2^n}$;

ii) $tr_m^n(x^{2^m}) = tr_m^n(x)$ for any $x \in F_{2^n}$; and

iii) $tr_1^n(x) = tr_1^m(tr_m^n(x))$ for all $x \in F_{2^n}$.

Let $f(\underline{v})$ be a function defined on the binary $n$-dimensional vector space $V_2^n$. $f(\underline{v})$ is called a *Boolean function* if it takes values in $\{0, 1\}$. The *Fourier transform* $f^w(\cdot)$ of a Boolean function $f(\underline{v})$ is defined by

$$f^w(\underline{\lambda}) = \sum_{\underline{v} \in V_2^n} (-1)^{f(\underline{v}) + \underline{\lambda} \cdot \underline{v}} \text{ for } \underline{\lambda} \in V_2^n \tag{11}$$

where $\underline{\lambda} \cdot \underline{v}$ denotes the inner product of two vectors $\underline{\lambda}$ and $\underline{v}$.

The *trace transform* of functions defined on $F_{2^n}$ was introduced by Olsen, Scholtz and Welch [15]. Let $g(x)$ be a function from $F_{2^n}$ to $F_2$. Its *trace transform* $G(\cdot)$ is defined by

$$G(\lambda) = \sum_{x \in F_{2^n}} (-1)^{g(x) + tr_1^n(\lambda x)} \text{ for } \lambda \in F_{2^n}. \tag{12}$$

By choosing two dual bases of $F_{2^n}$ over $F_2$ for $x$ and $\lambda$, respectively, then every function from $F_{2^n}$ to $F_2$, called Boolean function on $F_{2^n}$, can be expressed as a Boolean function on $V_2^n$, and the trace transform of a Boolean function on $F_{2^n}$ can be repressed as the Fourier transform of the associated Boolean function on $V_2^n$.

A Boolean function $f(x)$ on $F_{2^n}$ is a *quadratic form* if it can be written as a homogeneous polynomial of degree 2 on $V_2^n$, namely of the form $f(x_1, \cdots, x_n) = \sum\limits_{1 \le i < j \le n} a_{ij} x_i x_j$. The distribution of its Fourier transform values is completely determined by the rank of the symmetric matrix with zero diagonal entries and with $a_{ij}$ as the $(i, j)$ entry, or equivalently, determined by the number of the solutions on $z$ of the sympletic form

$$B_f(x, z) = f(x) + f(z) + f(x + z) = 0$$

holds for all $x$ in $F_{2^n}$. This rank is called the rank of $f(x)$. It must be even, and if it is $2h$ then the number of the solutions on $z$ is $2^{n-2h}$ [2].

*Theorem 1 ([2]):* Let $f(x)$ be a quadratic form on $F_{2^n}$ with rank $2h$, $1 \le h \le n/2$. Then its Fourier transform (trace transform) values has a distribution as

$$f^w(\lambda) = \begin{cases} \pm 2^{n-h}, & 2^{2h-1} \pm 2^{h-1} \text{ times}; \\ 0, & 2^n - 2^{2h} \text{ times}. \end{cases}$$

The following lemmas are simple facts from number theory and linear algebra. They will be used to prove results in this paper.

*Lemma 2:* Let $d = \gcd(n, k)$. Then $\gcd(2^n - 1, 2^k - 1) = 2^d - 1$ and

$$\gcd(2^n - 1, 2^k + 1) = \begin{cases} 1, & \text{if } n/d \text{ is odd}; \\ 1 + 2^d, & \text{otherwise}. \end{cases}$$

A *linearized polynomial* $L(x) \in F_{2^n}[x]$ is a polynomial of the form $L(x) = \sum\limits_{i=0}^{n-1} a_i x^{2^i}$.

*Lemma 3:* Let $L(x)$ be a linearized polynomial over $F_{2^n}$ and $\lambda \in F_{2^n}$. The equation $L(x) = \lambda$ has either no solution in $F_{2^n}$ or exactly the same number of solutions as the equation $L(x) = 0$.

*Lemma 4:* Let $\lambda$ be a nonzero element in the field $F_{2^n}$. If the equation

$$x^{2^k - 1} = \lambda$$

has a solution in $F_{2^n}$, then it has exactly $2^{\gcd(k,n)} - 1$ solutions in $F_{2^n}$.

For convenience, the following notations are used in the rest of this paper:

- $\mathbf{E} = F_{2^n}$: the finite field of $2^n$ elements;

- $\mathbf{E}^*$: the multiplicative group of $\mathbf{E}$;

- $\mathbf{F} = F_{2^{n/2}}$: the finite field of $2^{n/2}$ elements. (We always assume $n$ is even.) It is a subfield of $\mathbf{E}$;

- $\mathbf{F}^*$: the multiplicative group of $\mathbf{F}$;

- $\mathbf{E}^t = \mathbf{E} \times \mathbf{E} \times \cdots \times \mathbf{E}$: the Cartesian product of $t$ copies of $\mathbf{E}$;

- $\mathbf{E} \times \mathbf{F}$: the Cartesian product of $\mathbf{E}$ and $\mathbf{F}$;

- $\alpha$: a fixed primitive element of $\mathbf{E}$;

- $\beta = \alpha^{2^{n/2}+1}$: a primitive element of $\mathbf{F}$;

- $|\Phi|$: the cardinality of a set $\Phi$.

For any $b \in \mathbf{E}$ and $c \in \mathbf{F}$, a Boolean function $f_{b,c}(x)$ is defined by

$$f_{b,c}(x) = tr_1^n(bx^{2^k+1}) + tr_1^{n/2}(cx^{2^{n/2}+1}). \tag{13}$$

$f_{b,c}(x)$ is a quadratic form. For $b = 0$ or $c = 0$, its rank is determined in the following two propositions.

*Proposition 5:* Assume $b \in \mathbf{E}^*$ and $c = 0$.

(1) Assume $n \equiv 2 \bmod 4$ and $\gcd(k,n) = 2$. Then the rank of $f_{b,0}(x)$ is $n - 2$. Furthermore, $f_{b,0}^w(0) = 0$, and the distribution of $f_{b,0}^w(1)$ as $b$ runs through all elements in $\mathbf{E}^*$, is given by

$$f_{b,0}^w(1) = \begin{cases} \pm 2^{n/2+1}, & 2^{n-3} \pm 2^{n/2-2} \text{ times;} \\ 0, & 2^n - 2^{n-2} - 1 \text{ times.} \end{cases} \tag{14}$$

(2) Assume $n \equiv 0 \bmod 4$ and $\gcd(k,n) = 1$. Then the rank of $f_{b,0}(x)$ is $n - 2$ if $b$ is a cubic element in $\mathbf{E}$. Otherwise, the rank is $n$. When $b$ runs through all elements in $\mathbf{E}^*$, the distribution of $f_{b,0}^w(0)$ is

$$f_{b,0}^w(0) = \begin{cases} -2^{n/2+1}, & (2^n - 1)/3 \text{ times;} \\ 2^{n/2}, & 2(2^n - 1)/3 \text{ times} \end{cases} \tag{15}$$

and the distribution of $f_{b,0}^w(1)$ is

$$f_{b,0}^w(1) = \begin{cases} 2^{n/2+1}, & (2^{n-3} + 2^{n/2-2})/3 \text{ times}; \\ -2^{n/2+1}, & (2^{n-3} - 2^{n/2-2} - 1)/3 \text{ times}; \\ 0, & (2^n - 2^{n-2})/3 \text{ times}; \\ 2^{n/2}, & 2(2^{n-1} + 2^{n/2-1} - 1)/3 \text{ times}; \\ -2^{n/2}, & 2(2^{n-1} - 2^{n/2-1})/3 \text{ times}. \end{cases} \tag{16}$$

*Proof:* In the following, we assume $k < n/2$. The case of $k \geq n/2$ can be similarly proved. It is true that

$$B_{f_{b,0}}(x, z) = f_{b,0}(x) + f_{b,0}(z) + f_{b,0}(x+z) = tr_1^n[x(bz^{2^k} + b^{2^{n-k}}z^{2^{n-k}})].$$

The equation $B_{f_{b,0}}(x, z) = 0$ holds for all $x \in \mathbf{E}$ if and only if $bz^{2^k} + b^{2^{n-k}}z^{2^{n-k}} = 0$, or equivalently, if and only if $z = 0$ or $z^{2^k(2^{n-2k}-1)} = b^{1-2^{n-k}}$.

(1) Since $\gcd(k, n) = 2$ and $n/2$ is odd, one has $\gcd(n - 2k, n) = \gcd(k, n) = 2$. By Lemma 4, $z^{2^k(2^{n-2k}-1)} = b^{1-2^{n-k}}$ has 3 nonzero solutions on $z$ in $\mathbf{E}$. Thus, there are totally 4 solutions on $z$ such that $B_{f_{b,0}}(x, z) = 0$ holds for all $x \in \mathbf{E}$. Therefore, the rank of $f_{b,0}(x)$ is $n - 2$.

Since $\gcd(2^k + 1, 2^n - 1) = 1$ by Lemma 2, the function $bx^{2^k+1}$ is one-to-one on $\mathbf{E}$. Hence, $f_{b,0}^w(0) = 0$.

For any $a \in \mathbf{E}^*$, one has

$$f_{ba^{2^k+1},0}^w(a) = \sum_{x \in \mathbf{E}} (-1)^{tr_1^n(ax + ba^{2^k+1}x^{2^k+1})} = \sum_{y \in \mathbf{E}} (-1)^{tr_1^n(y + by^{2^k+1})} = f_{b,0}^w(1).$$

Therefore, for any fixed $a \in \mathbf{E}^*$, $f_{b,0}^w(a)$ and $f_{b,0}^w(1)$ has the same distribution when $b$ runs through all elements in $\mathbf{E}^*$. By Theorem 1, the distribution of $f_{b,0}^w(1)$ can be obtained as Eq. (14) when $b$ runs through all elements in $\mathbf{E}^*$.

(2) Since $n/2$ is even and $\gcd(k, n) = 1$, one has $\gcd(n - 2k, n) = 2$ and $\gcd(n - k, n) = 1$. By Lemma 2, $\gcd(2^{n-2k} - 1, 2^n - 1) = 3$ and $\gcd(2^{n-k} - 1, 2^n - 1) = 1$. The equation $z^{2^k(2^{n-2k}-1)} = b^{1-2^{n-k}}$ has exactly 3 solutions in $\mathbf{E}$ if $b$ is a cubic element in $\mathbf{E}^*$, and has no solution otherwise. Thus, the rank of $f_{b,c}(x)$ is $n - 2$ or $n$, depending on $b$ being a cubic element in $\mathbf{E}^*$ or not.

By Lemma 2 again, $\gcd(2^k + 1, 2^n - 1) = 3$. Thus, for any $b \in E^*$, the function $bx^{2^k+1}$ is three-to-one on $\mathbf{E}^*$. One has

$$\sum_{x \in \mathbf{E}^*} (-1)^{f_{b,0}(x)} \equiv 0 \bmod 3, \ f_{b,0}^w(0) \equiv 1 \bmod 3. \tag{17}$$

 

Since the rank of $f_{b,0}(x)$ is $n$ or $n-2$, by Theorem 1, $f_{b,0}^w(0) = 0, \pm 2^{n/2}$ or $\pm 2^{n/2+1}$. Thus, by Eq. (17), $f_{b,0}^w(0) = -2^{n/2+1}$ or $2^{n/2}$. Since

$$
\begin{aligned}
\sum_{b \in \mathbf{E}^*} f_{b,0}^w(0) &= \sum_{b \in \mathbf{E}^*} \sum_{x \in \mathbf{E}} (-1)^{tr_1^n(bx^{2^k+1})} \\
&= \sum_{x \in \mathbf{E}} \sum_{b \in \mathbf{E}^*} (-1)^{tr_1^n(bx^{2^k+1})} \\
&= 2^n - 1 + \sum_{x \in \mathbf{E}^*} \sum_{b \in \mathbf{E}^*} (-1)^{tr_1^n(bx^{2^k+1})} \\
&= 2^n - 1 + \sum_{x \in \mathbf{E}^*} (-1) = 0,
\end{aligned}
\tag{18}
$$

there are $(2^n - 1)/3$ and $2(2^n - 1)/3$ elements $b \in \mathbf{E}^*$ such that $f_{b,0}^w(0) = -2^{n/2+1}$ and $2^{n/2}$, respectively. Hence, the distribution of $f_{b,0}^w(1)$ can be obtained as Eq. (16). ∎

The following Proposition 6 can be proved in a similar way.

*Proposition 6:* Assume $c \in \mathbf{F}^*$ and $b = 0$. Then the rank of $f_{0,c}(x)$ is $n$. When $c$ runs through all elements in $\mathbf{F}^*$, the distribution of $f_{0,c}^w(0)$ is

$$
f_{0,c}^w(0) = \begin{cases} 2^{n/2}, & 0 \text{ times;} \\ -2^{n/2}, & 2^{n/2} - 1 \text{ times} \end{cases}
\tag{19}
$$

and the distribution of $f_{0,c}^w(1)$ is

$$
f_{0,c}^w(1) = \begin{cases} 2^{n/2}, & 2^{n/2-1} \text{ times;} \\ -2^{n/2}, & 2^{n/2-1} - 1 \text{ times.} \end{cases}
\tag{20}
$$

Proposition 6 can be applied to determine the correlation values of the small set of Kasami sequences and their distribution. The small set of Kasami sequences is a family of $2^{n/2}$ binary sequences of period $2^n - 1$ ($n \geq 4$ is even), defined by

$$
\mathcal{K}_s = \{\{tr_1^n(\alpha^t) + tr_1^{n/2}(\eta \alpha^{t(2^{n/2}+1)})\}_{t=0}^{2^n-2} \mid \eta \in \mathbf{F}\}.
\tag{21}
$$

It has optimal correlation property with respect to the Welch bound [5], [16].

Modifying the power index in Eq. (21), a family of sequences is defined by

$$
\{\{tr_1^n(\alpha^t) + tr_1^{n/2}(\eta \alpha^{t(2^k+1)})\}_{t=0}^{2^n-2} \mid \eta \in \mathbf{F}\}.
$$

When $n \equiv 2 \bmod 4$ and $\gcd(k, n) = 2$, by Proposition 5, the family has three valued out-of-phase auto- and cross-correlation values $-1$ and $\pm 2^{n/2+1} - 1$ [2].

There are several generalizations of the small set of Kasami sequences, including No sequences [17], TN sequences [18] and the generalized Kasami signal set [19]. These generalized sequence sets have the same correlation properties and family sizes as $\mathcal{K}_s$.

The large set of Kasami sequences has a large family size and contains the small set of Kasami sequences as its subset [5]. For even $n \geq 4$, define sequences $\{s_{\gamma\delta}(t)\}_{t=0}^{2^n-2}$ and $\{s_{\zeta\eta}(t)\}_{t=0}^{2^n-2}$ by

$$s_{\gamma\delta}(t) = tr_1^n(\alpha^t + \gamma\alpha^{t(2^{n/2+1}+1)}) + tr_1^{n/2}(\delta\alpha^{t(2^{n/2}+1)}) \tag{22}$$

and

$$s_{\zeta\eta}(t) = tr_1^n(\zeta\alpha^{t(2^{n/2+1}+1)}) + tr_1^{n/2}(\eta\alpha^{t(2^{n/2}+1)}). \tag{23}$$

The large set of Kasami sequences is

$$\mathcal{K}_l = \{\{s_{\gamma\delta}(t)\}_{t=0}^{2^n-2} \,|\, \gamma \in \mathbf{E}, \delta \in \mathbf{F}\} \bigcup \{\{s_{\zeta\eta}(t)\}_{t=0}^{2^n-2} \,|\, \zeta \in \Gamma, \ \eta \in \Delta\}, \tag{24}$$

where

$$\Gamma = \{1\}, \ \ \Delta = \mathbf{F} \tag{25}$$

for $n \equiv 2 \pmod 4$ and

$$\Gamma = \{1, \alpha, \alpha^2\}, \ \ \Delta = \{1, \beta, \cdots, \beta^{(2^{n/2}-1)/3-1}\} \tag{26}$$

for $n \equiv 0 \pmod 4$.

To determine the correlation distribution of $\mathcal{K}_l$, the rank of $f_{b,c}(x,z)$ with $b \neq 0$ and $c \neq 0$ is analyzed in the following sections.

### III. Some Equations over Finite Fields

This section studies the solutions of some equations over $\mathbf{E}$, which will be used in Section IV to determine the Fourier transform of $f_{b,c}(x)$.

A representation of elements in $\mathbf{E}$ is given as follows.

*Lemma 7:* Let $\beta$ and $\rho$ be two different elements in $\mathbf{E}$. Then

$$\mathbf{E} \setminus \{\beta\} = \left\{ \frac{x\beta + \rho^2}{x + \beta} \,\middle|\, x \in \mathbf{E} \setminus \{\beta\} \right\}. \tag{27}$$

*Proof:* Assume that

$$\frac{x_1\beta + \rho^2}{x_1 + \beta} = \frac{x_2\beta + \rho^2}{x_2 + \beta}$$

holds for $x_1, x_2 \in \mathbf{E} \setminus \{\beta\}$. Multiplying the both sides by $(x_1 + \beta)(x_2 + \beta)$, one can deduce that

$$(x_1 + x_2)(\beta^2 + \rho^2) = 0.$$

Since $\beta^2 + \rho^2 \neq 0$, one has $x_1 = x_2$. This shows that there are $2^n - 1$ different elements of the form $\frac{x\beta + \rho^2}{x + \beta}$. They are clearly different from $\beta$. Therefore Eq. (27) holds. ∎

The original version of Lemma 7 was proposed in [20] and [21] to describe subsets of the set

$$\mathcal{U} = \{x \in F_{2^n} \mid x^{2^{n/2}+1} = 1\}.$$

Lemma 7 enables us to estimate the number of solutions to the following Eq. (28).

*Proposition 8:* Assume $\gcd(l, n) = 1$ and $\varepsilon, \theta \in \mathbf{E}^*$. For any $v \in \mathbf{E}$, the equation

$$\varepsilon x^{2^l + 1} + vx + \theta = 0 \tag{28}$$

has at most 3 solutions in $\mathbf{E}$.

*Proof:* Assume Eq. (28) has a root $\beta$ in $\mathbf{E}$. Then $\beta \neq 0$. If $x \neq \beta$ is another solution to Eq (28), by Lemma 7, $x$ can be written as $u\beta/(u+\beta)$ for some $u \in \mathbf{E} \setminus \{\beta\}$. That is, all solutions to Eq. (28) different from $\beta$ are determined by the solutions to the equation

$$\varepsilon \left( \frac{u\beta}{u + \beta} \right)^{2^l + 1} + v \cdot \frac{u\beta}{u + \beta} + \theta = 0. \tag{29}$$

Multiplying both sides of Eq. (29) by $(u + \beta)^{2^l + 1}$ and simplifying the expansion, one has

$$\theta\beta u^{2^l} + (v\beta^{2^l + 1} + \theta\beta^{2^l})u + \theta\beta^{2^l + 1} = 0. \tag{30}$$

Since $\theta\beta \neq 0$, Eq. (30) is a nonzero linearized polynomial on $u$, and it has either no solution in $\mathbf{E}$ or exactly the same number of solutions as the equation

$$\theta\beta u^{2^l} + (v\beta^{2^l + 1} + \theta\beta^{2^l})u = 0. \tag{31}$$

Eq. (31) has at most two solutions, that is, it has at most one nonzero solution, since $\gcd(2^l - 1, 2^n - 1) = 1$ by Lemma 2. Therefore, Eq. (28) has at most three solutions. ∎

Proposition 8 can be applied to estimate the rank of $f_{b,c}(x)$ defined by Eq. (13) for $b \neq 0$ and $c \neq 0$.

For any $c \in \mathbf{F}^*$, there is an $e \in \mathbf{F}^*$ such that $c = e^{2^{n/2}+1}$ since $\gcd(2^{n/2} + 1, 2^{n/2} - 1) = 1$. Then,

$$f_{b,c}(e^{-1}x) = tr_1^n(be^{-2^k - 1}x^{2^k + 1}) + tr_1^{n/2}(x^{2^{n/2}+1}) = f_{\theta,1}(x),$$

where $\theta = be^{-2^k-1}$. Thus, $f_{b,c}(x)$ and $f_{\theta,1}(x)$ have the same rank. Since

$$
\begin{aligned}
B_{f_{\theta,1}}(x,z) &= tr_1^n(\theta x^{2^k+1}) + tr_1^{n/2}(x^{2^{n/2}+1}) + tr_1^n(\theta z^{2^k+1}) \\
&\quad + tr_1^{n/2}(z^{2^{n/2}+1}) + tr_1^n[\theta(x+z)^{2^k+1}] + tr_1^{n/2}[(x+z)^{2^{n/2}+1}] \\
&= tr_1^n(\theta x^{2^k}z + \theta x z^{2^k}) + tr_1^{n/2}(x^{2^{n/2}}z + xz^{2^{n/2}}) \\
&= tr_1^n(\theta^{2^{n-k}}xz^{2^{n-k}} + \theta xz^{2^k}) + tr_1^{n/2}[tr_{n/2}^n(xz^{2^{n/2}})] \\
&= tr_1^n[x(\theta^{2^{n-k}}z^{2^{n-k}} + \theta z^{2^k} + z^{2^{n/2}})],
\end{aligned}
$$

the rank of $f_{\theta,1}(x)$ is determined by the number of solutions on $z$ to the equation

$$\theta^{2^{n-k}}z^{2^{n-k}} + \theta z^{2^k} + z^{2^{n/2}} = 0. \tag{32}$$

For any $k < n/2$ with $\gcd(n/2-k, n) = 1$, one has

$$\theta^{2^{n-k}}z^{2^{n-k}} + \theta z^{2^k} + z^{2^{n/2}} = (\theta^{2^{n-k}}w^{2^{n/2-k}+1} + w + \theta)z^{2^k} = 0$$

where $w = z^{2^k(2^{n/2-k}-1)}$. Since $\gcd(2^k(2^{n/2-k}-1), 2^n-1) = 1$, there is a one-to-one correspondence between elements $w$ and $z$. Thus, Eq. (32) has the same number of nonzero solutions as that for the equation

$$\theta^{2^{n-k}}w^{2^{n/2-k}+1} + w + \theta = 0. \tag{33}$$

Similarly, for any $k > n/2$ with $\gcd(k-n/2, n) = 1$, Eq. (32) and the equation

$$\theta w^{2^{k-n/2}+1} + w + \theta^{2^{n-k}} = 0 \tag{34}$$

have the same number of nonzero solutions.

Applying Proposition 8, Eq. (33) and Eq. (34) have at most three solutions. Thus, the rank of $f_{b,c}(x)$ is $n-2$ or $n$. This implies that the number of solutions to Eq. (32) is either $1$ or $4$. Thus, for any $(b,c) \in \mathbf{E}^* \times \mathbf{F}^*$, Eq. (32) and Eq. (34) have either no or exactly three solutions.

Note that for even $n$, the condition $\gcd(n/2-k, n) = 1$ is equivalent to the following condition:

$$\gcd(k,n) = 2 \text{ if } n/2 \text{ is odd, and } \gcd(k,n) = 1 \text{ if } n/2 \text{ is even.} \tag{35}$$

It is assumed in the sequel that Eq. (35) holds.

Let $N_1$ and $N_2$ be the numbers of $\theta \in \mathbf{E}^*$ such that Eq. (33) and Eq. (34) has three solutions, respectively. A technique of coding theory is used to determine $N_1$ and $N_2$ in the following.

Definition 9 generalizes the concept of Kasami code [22].

*Definition 9:* For $\gamma, \delta \in \mathbf{E}$ and $\eta \in \mathbf{F}$, let $c(\gamma, \delta, \eta)$ be the $(2^n - 1)$-tuple vector indexed by the elements in $\mathbf{E}^*$ as

$$c(\gamma, \delta, \eta) = (tr_1^n(\gamma x + \delta x^{2^k+1}) + tr_1^{n/2}(\eta x^{2^{n/2}+1}), x \in \mathbf{E}^*). \tag{36}$$

The *generalized Kasami code* is the $[2^n - 1, 5n/2]$ code defined by

$$\mathcal{C} = \{c(\gamma, \delta, \eta) \mid \gamma, \delta \in \mathbf{E}, \eta \in \mathbf{F}\}. \tag{37}$$

Denote by $\mathcal{D}$ the dual code of $\mathcal{C}$. Let $A_j$ and $B_j$ denote the numbers of vectors of weight $j$ in $\mathcal{C}$ and $\mathcal{D}$, respectively.

*Proposition 10:* $B_1 = B_2 = B_3 = 0$.

*Proof:* We prove $B_3 = 0$. $B_1 = B_2 = 0$ can be more simply proved.

If $B_3 > 0$ then there are integers $0 \leq i_1 < i_2 < i_3 \leq 2^n - 2$ such that

$$\sum_{j=1}^{3}[tr_1^n(\gamma \alpha^{i_j} + \delta \alpha^{i_j(2^k+1)}) + tr_1^{n/2}(\eta \alpha^{i_j(2^{n/2}+1)})] = 0$$

holds for any $\gamma, \delta \in \mathbf{E}$ and $\eta \in \mathbf{F}$.

Take $\delta = \eta = 0$, one has

$$tr_1^n[\gamma(\alpha^{i_1} + \alpha^{i_2} + \alpha^{i_3})] = 0$$

holds for any $\gamma \in \mathbf{E}$. Thus,

$$\alpha^{i_1} + \alpha^{i_2} + \alpha^{i_3} = 0. \tag{38}$$

Similarly, one has

$$\alpha^{i_1(2^k+1)} + \alpha^{i_2(2^k+1)} + \alpha^{i_3(2^k+1)} = 0 \tag{39}$$

and

$$\alpha^{i_1(2^{n/2}+1)} + \alpha^{i_2(2^{n/2}+1)} + \alpha^{i_3(2^{n/2}+1)} = 0. \tag{40}$$

By Eq. (38), one has

$$\alpha^{i_1(2^k+1)} = (\alpha^{i_2} + \alpha^{i_3})^{2^k+1} = \alpha^{i_2(2^k+1)} + \alpha^{i_2 2^k + i_3} + \alpha^{i_3 2^k + i_2} + \alpha^{i_3(2^k+1)}. \tag{41}$$

Then Eq. (39) and Eq. (41) imply

$$(i_2 - i_3)(2^k - 1) \equiv 0 \bmod (2^n - 1).$$

Similarly, Eq. (38) and Eq. (40) imply

$$(i_2 - i_3)(2^{n/2} - 1) \equiv 0 \bmod (2^n - 1).$$

Since $\gcd(k, n/2) = 1$, one has $\gcd(2^k - 1, 2^{n/2} - 1) = 1$. There are integers $t_1, t_2$ such that

$$1 = t_1(2^k - 1) + t_2(2^{n/2} - 1).$$

Thus,

$$i_2 - i_3 \equiv t_1(i_2 - i_3)(2^k - 1) + t_2(i_2 - i_3)(2^{n/2} - 1) \equiv 0 \bmod (2^n - 1).$$

This shows $i_2 = i_3$, which contradicts the assumption $i_2 < i_3$. Thus, $B_3 = 0$. ∎

In the following theorem, the weight enumerator of $\mathcal{C}$ is provided in terms of $N_1$ or $N_2$. Then, Proposition 10 is applied to determine the values of $N_1$ and $N_2$ and find the weight distribution of $\mathcal{C}$.

*Theorem 11:* (1)

$$N_1 = N_2 = \begin{cases} (2^{n+1} - 2^{n/2+1} - 4)/3, & \text{if } n/2 \text{ is odd;} \\ (2^{n+1} - 2)/3, & \text{if } n/2 \text{ is even;} \end{cases}$$

(2) The code $\mathcal{C}$ has the following weight distribution:

$$\begin{cases} A_{2^{n-1} \pm 2^{n/2}} = (2^{n/2} - 1)(2^{n-3} \mp 2^{n/2-2})(2^{n+1} + 2^{n/2} - 1)/3; \\ A_{2^{n-1}} = (2^{n/2} - 1)(2^{2n-1} + 2^{3n/2-2} - 2^{n-2} + 2^{n/2} + 1); \\ A_{2^{n-1} \pm 2^{n/2-1}} = (2^{n/2} - 1)(2^{n-1} \mp 2^{n/2-1})(2^n + 2^{n/2+1} + 4)/3. \end{cases} \tag{42}$$

*Proof:* Assume $k < n/2$. The proof for the case of $k > n/2$ is similar.

Let

$$g(x) = tr_1^n(\gamma x + \delta x^{2^k+1}) + tr_1^{n/2}(\eta x^{2^{n/2}+1}).$$

The proof is divided into four cases as follows.

*Case 1:* $\eta = \delta = 0$ and $\gamma \neq 0$

The function $g(x)$ is a linear function from $\mathbf{E}$ to $F_2$, and the weight of $c(\gamma, \delta, \eta)$ is $2^{n-1}$.

*Case 2:* $\eta = 0$ and $\delta \neq 0$

When $n/2$ is odd, by Proposition 5 (1), the rank of $tr_1^n(\delta x^{2^k+1})$ is $n - 2$. By Theorem 1, the weight distribution of $c(\gamma, \delta, \eta)$ as $\gamma$ runs through all elements in $\mathbf{E}$ is given by

$$\begin{cases} 2^{n-1} \pm 2^{n/2} & \text{occuring } 2^{n-3} \mp 2^{n/2-2} \text{ times;} \\ 2^{n-1} & \text{occuring } 2^n - 2^{n-2} \text{ times.} \end{cases}$$

When $n/2$ is even, by Proposition 5(2), the rank of $tr_1^n(\delta x^{2^k+1})$ is $n-2$ or $n$, depending on $\delta$ is a cubic element in $\mathbf{E}$ or not. By Theorem 1, the weight distribution of $c(\gamma,\delta,\eta)$ as $\gamma$ runs through all elements in $\mathbf{E}$ is

$$\begin{cases} 2^{n-1} \pm 2^{n/2} & \text{occuring } 2^{n-3} \mp 2^{n/2-2} \text{ times;} \\ 2^{n-1} & \text{occuring } 2^n - 2^{n-2} \text{ times} \end{cases}$$

or

$$2^{n-1} \pm 2^{n/2-1} \quad \text{occuring } 2^{n-1} \mp 2^{n/2-1} \text{ times}$$

depending on $\delta$ is a cubic element or not.

*Case 3:* $\eta \neq 0$ and $\delta = 0$

By Proposition 6, the rank of $tr_1^{n/2}(\eta x^{2^{n/2}+1})$ is $n$. By Theorem 1, the weight distribution of $c(\gamma,\delta,\eta)$ as $\gamma$ runs through all elements in $\mathbf{E}$, is

$$2^{n-1} \pm 2^{n/2-1} \quad \text{occuring } 2^{n-1} \mp 2^{n/2-1} \text{ times.}$$

*Case 4:* $\eta \neq 0$ and $\delta \neq 0$

By the analysis after Proposition 8, for any fixed nonzero $\eta$, the weight distribution of $c(\gamma,\delta,\eta)$ as $(\gamma,\delta)$ runs through all elements in $\mathbf{E} \times \mathbf{E}^*$, is

$$\begin{cases} 2^{n-1} \pm 2^{n/2} & \text{occuring } N_1(2^{n-3} \mp 2^{n/2-2}) \text{ times;} \\ 2^{n-1} & \text{occuring } N_1(2^n - 2^{n-2}) \text{ times;} \\ 2^{n-1} \pm 2^{n/2-1} & \text{occuring } (2^n - 1 - N_1)(2^n \mp 2^{n-2}) \text{ times.} \end{cases}$$

Combining the results for all four cases, the weight distribution of $\mathcal{C}$ is

$$\begin{cases} A_{2^{n-1}\pm 2^{n/2}} = (2^{n/2}-1)(2^{n-3} \mp 2^{n/2-2})(2^{n/2}+1+N_1); \\ A_{2^{n-1}} = (2^{n/2}-1)[(2^n-2^{n-2}+1)(2^{n/2}+1)+N_1(2^n-2^{n-2})]; \\ A_{2^{n-1}\pm 2^{n/2-1}} = (2^{n/2}-1)(2^{n-1} \mp 2^{n/2-1})(2^n - N_1) \end{cases}$$

for $n/2$ odd, and the weight distribution is

$$\begin{cases} A_{2^{n-1}\pm 2^{n/2}} = (2^{n/2}-1)(2^{n-3} \mp 2^{n/2-2})(2^{n/2}+1+3N_1); \\ A_{2^{n-1}} = 2^n-1+(2^{n/2}-1)(2^n-2^{n-2})(2^{n/2}+1+3N_1)/3; \\ A_{2^{n-1}\pm 2^{n/2-1}} = (2^{n/2}-1)(2^{n-1} \mp 2^{n/2-1})(3 \cdot 2^n + 2^{n/2+1}+2-3 \cdot N_1)/3 \end{cases}$$

for $n/2$ even. Therefore, one has

$$A_{2^{n-1}+2^{n/2}} + A_{2^{n-1}-2^{n/2}} + A_{2^{n-1}} + A_{2^{n-1}+2^{n/2-1}} + A_{2^{n-1}-2^{n/2-1}} = 2^{5n/2}-1, \quad (43)$$

$$2(A_{2^{n-1}-2^{n/2}} - A_{2^{n-1}+2^{n/2}}) + A_{2^{n-1}-2^{n/2-1}} - A_{2^{n-1}+2^{n/2-1}} = 2^{2n} - 2^{n/2}, \tag{44}$$

and

$$4(A_{2^{n-1}-2^{n/2}} + A_{2^{n-1}+2^{n/2}}) + A_{2^{n-1}-2^{n/2-1}} + A_{2^{n-1}+2^{n/2-1}} = 2^{5n/2} - 2^n. \tag{45}$$

The weight enumerator of $\mathcal{C}$ is given by

$$
\begin{aligned}
W_{\mathcal{C}}(x, y) \;=\;& A_{2^{n-1}+2^{n/2}} x^{2^{n-1}-2^{n/2}-1} y^{2^{n-1}+2^{n/2}} + A_{2^{n-1}-2^{n/2}} x^{2^{n-1}+2^{n/2}-1} y^{2^{n-1}-2^{n/2}} \\
&+ A_{2^{n-1}} x^{2^{n-1}-1} y^{2^{n-1}} + A_{2^{n-1}+2^{n/2-1}} x^{2^{n-1}-2^{n/2-1}-1} y^{2^{n-1}+2^{n/2-1}} \\
&+ A_{2^{n-1}-2^{n/2-1}} x^{2^{n-1}+2^{n/2-1}-1} y^{2^{n-1}-2^{n/2-1}} + x^{2^n-1}.
\end{aligned} \tag{46}
$$

By Eq. (10), the weight enumerator of $\mathcal{D}$ is given by

$$W_{\mathcal{D}}(x, y) = 2^{-5n/2} W_{\mathcal{C}}(x + y, x - y). \tag{47}$$

Then $2^{5n/2} B_3$ is the coefficient of $x^{2^n-4} y^3$ in the expansion of $W_{\mathcal{C}}(x + y, x - y)$. Utilizing Eq. (43), Eq. (44), and Eq. (45), one has

$$
\begin{aligned}
6 \cdot 2^{5n/2} \cdot B_3 = 2^{3n/2}[8(A_{2^{n-1}-2^{n/2}} - A_{2^{n-1}+2^{n/2}}) + A_{2^{n-1}-2^{n/2-1}} - A_{2^{n-1}+2^{n/2-1}}] \\
+ 2^{5n/2+1} - 3 \cdot 2^{7n/2} + 2^{3n}.
\end{aligned}
$$

Therefore,

$$6 \cdot 2^{5n/2} \cdot B_3 = 2^{2n}(2^{n/2} - 1)(2^n + 3N_1 + 2^{n/2+2} + 4) + 2^{5n/2+1} - 3 \cdot 2^{7n/2} + 2^{3n} \tag{48}$$

for odd $n/2$, and

$$6 \cdot 2^{5n/2} \cdot B_3 = 2^{2n}(2^{n/2} - 1)(2^n + 3N_1 + 2^{n/2+1} + 2) + 2^{5n/2+1} - 3 \cdot 2^{7n/2} + 2^{3n} \tag{49}$$

for even $n/2$.

By Proposition 10, $B_3 = 0$. From Eq. (48) and Eq. (49), the value of $N_1$ can be determined as in Theorem 11 (1). Furthermore, the weight distribution is deduced as in Eq. (42). ∎

The weight distribution of $\mathcal{C}$ is the same as that of the code in Theorem 14 of [11]. From Theorem 11 and the analysis after Proposition 8, the following Corollary 12 is obtained. The corollary will be used to determine the Fourier transform values of $f_{b,c}(x)$ in the next section.

*Corollary 12:* For any $c \in \mathbf{F}^*$, the rank of $f_{b,c}(x)$ is either $n - 2$ or $n$. Further, if $n/2$ is odd then there are exactly $(2^{n+1} - 2^{n/2+1} - 4)/3$ values of $b \in \mathbf{E}^*$ such that the rank of $f_{b,c}(x)$ is $n - 2$; if $n/2$ is even, there are exactly $(2^{n+1} - 2)/3$ values of $b$ with the same property.

## IV. FOURIER TRANSFORM OF $f_{b,c}(x)$

Combining Corollary 12, Theorem 1, and Propositions 5 and 6, one has

*Proposition 13:* The distribution of Fourier transform values of functions $f_{b,c}(x)$ as $(b, c)$ runs through all elements in $\mathbf{E} \times \mathbf{F}$ is given by

$$
\begin{cases}
\pm 2^{n/2+1}, & (2^{n/2} - 1)(2^{n-3} \pm 2^{n/2-2})(2^{n+1} + 2^{n/2} - 1)/3 \text{ times}; \\
0, & (2^{n/2} - 1)(2^{2n-1} + 2^{3n/2-2} - 2^{n-2} + 2^{n/2} + 1) \text{ times}; \\
\pm 2^{n/2}, & (2^{n/2} - 1)(2^{n-1} \pm 2^{n/2-1})(2^{n} + 2^{n/2+1} + 4)/3 \text{ times}; \\
2^{n}, & 1 \text{ time}.
\end{cases}
\tag{50}
$$

In order to completely determine the correlation distribution of the families proposed in Section V, the distributions on $f_{b,c}^{w}(0)$ and $f_{b,c}^{w}(1)$ are further considered. These distributions are closely related to the following sets:

$$
\Pi_1 = \{(x, y, z) \in \mathbf{E}^3 \mid x^{2^k+1} + y^{2^k+1} + z^{2^k+1} = 0\}
\tag{51}
$$

$$
\Pi_2 = \{(x, y, z) \in \mathbf{E}^3 \mid x^{2^{n/2}+1} + y^{2^{n/2}+1} + z^{2^{n/2}+1} = 0\}.
\tag{52}
$$

Their cardinalities are as follows.

*Lemma 14:*

$$
|\Pi_1| =
\begin{cases}
2^{2n}, & \text{if } n/2 \text{ is odd}; \\
2^{2n} - 2^{3n/2+1} + 2^{n/2+1}, & \text{if } n/2 \text{ is even};
\end{cases}
$$

and

$$
|\Pi_2| = 2^{5n/2} - 2^{3n/2} + 2^{n}.
$$

*Proof:* By Proposition 5, one has

$$
\sum_{b \in \mathbf{E}^*} [f_{b,0}^{w}(0)]^3 =
\begin{cases}
0, & \text{if } n/2 \text{ is odd}; \\
-2^{3n/2+1}(2^{n} - 1), & \text{if } n/2 \text{ is even}.
\end{cases}
$$

On the other hand, one has

$$
\begin{aligned}
\sum_{b \in \mathbf{E}^*} [f_{b,0}^{w}(0)]^3 &= \sum_{b \in \mathbf{E}^*} \sum_{x \in \mathbf{E}} (-1)^{tr_1^n(bx^{2^k+1})} \sum_{y \in \mathbf{E}} (-1)^{tr_1^n(by^{2^k+1})} \sum_{z \in \mathbf{E}} (-1)^{tr_1^n(bz^{2^k+1})} \\
&= \sum_{b \in \mathbf{E}^*} \sum_{x,y,z \in \mathbf{E}} (-1)^{tr_1^n[b(x^{2^k+1} + y^{2^k+1} + z^{2^k+1})]} \\
&= \sum_{x,y,z \in \mathbf{E}} \sum_{b \in \mathbf{E}^*} (-1)^{tr_1^n[b(x^{2^k+1} + y^{2^k+1} + z^{2^k+1})]} \\
&= \sum_{(x,y,z) \in \Pi_1} (2^n - 1) + \sum_{(x,y,z) \notin \Pi_1} \sum_{b \in \mathbf{E}^*} (-1)^{tr_1^n[b(x^{2^k+1} + y^{2^k+1} + z^{2^k+1})]} \\
&= (2^n - 1)|\Pi_1| + (2^{3n} - |\Pi_1|)(-1) \\
&= 2^n |\Pi_1| - 2^{3n}.
\end{aligned}
$$

So, it is easy to get the value of $|\Pi_1|$.

The cardinality of $\Pi_2$ can be calculated as follows. For any $(x, y) \in \mathbf{E}^2$ with $x^{2^{n/2}+1} \neq y^{2^{n/2}+1}$, there are $2^{n/2} + 1$ values for $z$ such that $(x, y, z) \in \Pi_2$, since $x^{2^{n/2}+1}$ is $(2^{n/2}+1)$-to-one on $\mathbf{E}^*$ and $\gcd(2^{n/2} + 1, 2^{n/2} - 1) = 1$. When $(x, y) \in \mathbf{E}^2$ satisfies $x^{2^{n/2}+1} = y^{2^{n/2}+1}$, $z = 0$ must be true for each tuple $(x, y, z)$ in $\Pi_2$. There are $(2^n - 1)(2^{n/2} + 1)$ nonzero tuples $(x, y)$ to make $x^{2^{n/2}+1} = y^{2^{n/2}+1}$ holds. Thus,

$$
\begin{aligned}
|\Pi_2| &= [2^{2n} - (2^n - 1)(2^{n/2} + 1) - 1](2^{n/2} + 1) + (2^n - 1)(2^{n/2} + 1) + 1 \\
&= 2^{5n/2} - 2^{3n/2} + 2^n.
\end{aligned}
$$

∎

It is necessary to find the cardinality of $\Pi_1 \cap \Pi_2$.

*Lemma 15:*

$$
|\Pi_1 \cap \Pi_2| = 3 \cdot 2^n - 2.
$$

*Proof:* We show the solutions $(x, y, z) \in \mathbf{E}$ to the system of equations

$$
\begin{cases}
x^{2^k+1} + y^{2^k+1} + z^{2^k+1} = 0 \\
x^{2^{n/2}+1} + y^{2^{n/2}+1} + z^{2^{n/2}+1} = 0
\end{cases}
\tag{53}
$$

are given by

$$
\{(0, 0, 0), (x, x, 0), (x, 0, x), (0, x, x) \mid x \in \mathbf{E}^*\}.
$$

Let $(x, y, z)$ be a solution to Eq. (53). If $z = 0$, then

$$
\begin{cases}
x^{2^k+1} + y^{2^k+1} = 0; \\
x^{2^{n/2}+1} + y^{2^{n/2}+1} = 0.
\end{cases}
$$

Since $\gcd(k, n/2 - k) = 1$ and $n/2 - k$ is odd, by Lemma 2, one has

$$
\gcd(2^k + 1, 2^{n/2} + 1) = \gcd(2^k + 1, 2^{n/2} - 2^k) = \gcd(2^k + 1, 2^{n/2-k} - 1) = 1.
$$

By an approach similar to the proof of Proposition 10, one has $x = y$.

If $z \neq 0$, then one has

$$
\begin{cases}
x_1^{2^k+1} + y_1^{2^k+1} + 1 = 0 \\
x_1^{2^{n/2}+1} + y_1^{2^{n/2}+1} + 1 = 0
\end{cases}
$$

where $x_1 = x/z$ and $y_1 = y/z$. Thus,

$$
x_1^{(2^k+1)(2^{n/2}+1)} = (y_1^{2^k+1} + 1)^{(2^{n/2}+1)} = y_1^{(2^k+1)(2^{n/2}+1)} + y_1^{(2^k+1)2^{n/2}} + y_1^{2^k+1} + 1
\tag{54}
$$

and

$$x_1^{(2^{n/2}+1)(2^k+1)} = (y_1^{2^k+1} + 1)^{2^{n/2}+1} = y_1^{(2^{n/2}+1)(2^k+1)} + y_1^{(2^{n/2}+1)2^k} + y_1^{2^{n/2}+1} + 1. \quad (55)$$

From Eq. (54) and Eq. (55), one has

$$(y_1^{2^{n/2+k}} + y_1)(y_1^{2^{n/2}} + y_1^{2^k}) = 0,$$

which implies $y_1 = 0$ or $1$, since $\gcd(n/2 \pm k, n) = 1$. Thus, $(x_1, y_1)$ is equal to $(1, 0)$ or $(0, 1)$. Therefore, all solutions to Eq. (53) are obtained. ∎

The following power sums are calculated from cardinalities of $\Pi_1$, $\Pi_2$, and $\Pi_1 \cap \Pi_2$. They will be used in Proposition 17 to find the distributions on $f_{b,c}^w(0)$ and $f_{b,c}^w(1)$.

*Lemma 16:*

$$\sum_{(b,c) \in \mathbf{E}^* \times \mathbf{F}^*} f_{b,c}^w(0) = 2^{n/2}(2^n - 1);$$

$$\sum_{(b,c) \in \mathbf{E}^* \times \mathbf{F}^*} [f_{b,c}^w(0)]^2 = \begin{cases} 2^n(2^n - 1)(2^{n/2} - 1), & \text{for odd n/2;} \\ 2^n(2^n - 1)(2^{n/2} - 3), & \text{for even } n/2; \end{cases}$$

$$\sum_{(b,c) \in \mathbf{E}^* \times \mathbf{F}^*} [f_{b,c}^w(0)]^3 = \begin{cases} -2^{3n/2}(2^n - 1)(2^{n/2} - 3), & \text{for odd n/2;} \\ -2^{3n/2}(2^n - 1)(2^{n/2} - 5), & \text{for even } n/2. \end{cases}$$

*Proof:* The degree 1 power sum can be directly calculated.

With Lemmas 14 and 15, one has

$$
\begin{aligned}
&\sum_{(b,c) \in \mathbf{E}^* \times \mathbf{F}^*} (f_{b,c}^w(0))^3 \\
=~& \sum_{b \in \mathbf{E}^*} \sum_{c \in \mathbf{F}^*} \sum_{x,y,z \in \mathbf{E}} (-1)^{tr_1^n[b(x^{2^k+1}+y^{2^k+1}+z^{2^k+1})]+tr_1^{n/2}[c(x^{2^{n/2}+1}+y^{2^{n/2}+1}+z^{2^{n/2}+1})]} \\
=~& \sum_{x,y,z \in \mathbf{E}} \sum_{b \in \mathbf{E}^*} (-1)^{tr_1^n[b(x^{2^k+1}+y^{2^k+1}+z^{2^k+1})]} \sum_{c \in \mathbf{F}^*} (-1)^{tr_1^{n/2}[c(x^{2^{n/2}+1}+y^{2^{n/2}+1}+z^{2^{n/2}+1})]} \\
=~& \sum_{(x,y,z) \in \Pi_1 \cap \Pi_2} (2^n - 1)(2^{n/2} - 1) + \sum_{(x,y,z) \in \Pi_2 \setminus \Pi_1} (-1)(2^{n/2} - 1) + \sum_{(x,y,z) \in \Pi_1 \setminus \Pi_2} (-1)(2^n - 1) \\
&+ \sum_{(x,y,z) \notin \Pi_1 \cup \Pi_2} (-1)(-1) \\
=~& 2^{3n/2}|\Pi_1 \cap \Pi_2| - 2^{n/2}|\Pi_2| - 2^n|\Pi_1| + 2^{3n} \\
=~& \begin{cases} -2^{3n/2}(2^n - 1)(2^{n/2} - 3), & \text{for odd } n/2; \\ -2^{3n/2}(2^n - 1)(2^{n/2} - 5), & \text{for even } n/2. \end{cases}
\end{aligned}
$$

Let

$$\Phi_1 = \{(x,y) \in \mathbf{E}^2 \,|\, x^{2^k+1} = y^{2^k+1}\} \text{ and } \Phi_2 = \{(x,y) \in \mathbf{E}^2 \,|\, x^{2^{n/2}+1} = y^{2^{n/2}+1}\}.$$

Then by Lemma 15,

$$|\Phi_1 \cap \Phi_2| = |\{(x,y,0) \in \Pi_1 \cap \Pi_2\}| = |\{(x,x,0)|x \in E\}| = 2^n.$$

Note that

$$\gcd(2^k + 1, 2^n - 1) = \begin{cases} 1, & \text{for odd } n/2; \\ 3, & \text{for even } n/2 \end{cases}$$

and $(2^{n/2} + 1)|(2^n - 1)$, so,

$$|\Phi_1| = \begin{cases} 2^n, & \text{for odd } n/2; \\ 1 + 3(2^n - 1), & \text{for even } n/2 \end{cases}$$

and

$$|\Phi_2| = 1 + (2^{n/2} + 1)(2^n - 1).$$

Similarly as for the degree 3 power sum, one has

$$\sum_{(b,c) \in \mathbf{E}^* \times \mathbf{F}^*} (f_{b,c}^w(0))^2 = 2^{3n/2}|\Phi_1 \cap \Phi_2| - 2^{n/2}|\Phi_2| - 2^n|\Phi_1| + 2^{2n}$$

$$= \begin{cases} 2^n(2^n - 1)(2^{n/2} - 1), & \text{for } n/2 \text{ odd}; \\ 2^n(2^n - 1)(2^{n/2} - 3), & \text{for } n/2 \text{ even}. \end{cases}$$

■

*Proposition 17:* (1) For odd $n/2$, when $(b,c)$ runs through all elements in $\mathbf{E}^* \times \mathbf{F}^*$, the distribution of $f_{b,c}^w(0)$ is

$$f_{b,c}^w(0) = \begin{cases} 2^{n/2+1}, & 0 \text{ time}; \\ -2^{n/2+1}, & (2^n - 1)(2^{n/2-1} - 1)/3 \text{ times}; \\ 0, & (2^n - 1)(2^{n/2-1} - 1) \text{ times}; \\ 2^{n/2}, & (2^n - 1)(2^{n/2} + 1)/3 \text{ times}; \\ -2^{n/2}, & 0 \text{ time} \end{cases} \tag{56}$$

and the distribution of $f_{b,c}^w(1)$ is

$$f_{b,c}^w(1) = \begin{cases} 2^{n/2+1}, & (2^{n-3} + 2^{n/2-2})(2^{n/2+1} - 4)/3 \text{ times}; \\ -2^{n/2+1}, & (2^{3n/2-2} - 2^n + 2^{n/2-1} + 1)/3 \text{times}; \\ 0, & 2^{3n/2-1} - 2^n - 2^{n/2-1} + 1 \text{ times}; \\ 2^{n/2}, & (2^{n-1} + 2^{n/2-1} - 1)(2^{n/2} + 1)/3 \text{ times}; \\ -2^{n/2}, & (2^{n-1} - 2^{n/2-1})(2^{n/2} + 1)/3 \text{ times}. \end{cases} \tag{57}$$

(2) For even $n/2$, when $(b, c)$ runs through all elements in $\mathbf{E}^* \times \mathbf{F}^*$, the distribution of $f_{b,c}^w(0)$ is

$$
f_{b,c}^w(0) = \begin{cases}
2^{n/2+1}, & 0 \text{ time}; \\
-2^{n/2+1}, & (2^n - 1)(2^{n/2-1} - 2)/3 \text{ times}; \\
0, & (2^n - 1)2^{n/2-1} \text{ times}; \\
2^{n/2}, & (2^n - 1)(2^{n/2} - 1)/3 \text{ times}; \\
-2^{n/2}, & 0 \text{ time}
\end{cases}
\tag{58}
$$

and the distribution of $f_{b,c}^w(1)$ is

$$
f_{b,c}^w(1) = \begin{cases}
2^{n/2+1}, & (2^{n/2+1} - 2)(2^{n-3} + 2^{n/2-2})/3 \text{ times}; \\
-2^{n/2+1}, & (2^{3n/2-2} - 3 \cdot 2^{n-2} + 2)/3 \text{ times}; \\
0, & 2^{3n/2-1} - 2^{n-1} - 2^{n/2-1} \text{ times}; \\
2^{n/2}, & (2^{n-1} + 2^{n/2-1} - 1)(2^{n/2} - 1)/3 \text{ times}; \\
-2^{n/2}, & (2^{n-1} - 2^{n/2-1})(2^{n/2} - 1)/3 \text{ times}.
\end{cases}
\tag{59}
$$

*Proof:* (1) Let the distribution of $f_{b,c}^w(0)$ as $(b, c)$ runs through all elements in $\mathbf{E}^* \times \mathbf{F}^*$, be $2^{n/2+1}$, $x_1$ times; $-2^{n/2+1}$, $x_2$ times; $0$, $x_3$ times; $2^{n/2}$, $x_4$ times; $-2^{n/2}$, $x_5$ times.

From Corollary 12 and Lemma 16, one has

$$
\begin{cases}
x_1 + x_2 + x_3 = 2(2^n - 2^{n/2} - 2)(2^{n/2} - 1)/3; \\
x_4 + x_5 = (2^{n/2} + 1)^2(2^{n/2} - 1)/3; \\
2^{n/2+1}(x_1 - x_2) + 2^{n/2}(x_4 - x_5) = 2^{n/2}(2^n - 1); \\
(2^{n/2+1})^2(x_1 + x_2) + (2^{n/2})^2(x_4 + x_5) = 2^n(2^n - 1)(2^{n/2} - 1); \\
(2^{n/2+1})^3(x_1 - x_2) + (2^{n/2})^3(x_4 - x_5) = -2^{3n/2}(2^n - 1)(2^{n/2} - 3).
\end{cases}
\tag{60}
$$

The values of $x_i$ $(1 \leq i \leq 5)$ can be found by solving Eq. (60).

By the analysis after Proposition 8, the distribution of the Fourier transform values of functions $f_{b,c}(x)$ as $(b, c)$ runs through all elements in $\mathbf{E}^* \times \mathbf{F}^*$, is

$$
\begin{cases}
\pm 2^{n/2+1}, & (2^{n/2} - 1)(2^{n-3} \pm 2^{n/2-2})(2^{n+1} - 2^{n/2+1} - 4)/3 \text{ times} \\
0, & (2^{n/2} - 1)(2^n - 2^{n-2})(2^{n+1} - 2^{n/2+1} - 4)/3 \text{ times} \\
\pm 2^{n/2}, & (2^{n/2} - 1)(2^{n-1} \pm 2^{n/2-1})(2^n + 2^{n/2+1} + 1)/3 \text{ times}.
\end{cases}
$$

The approach used in the proof of Proposition 5(1) can also show that, for any $a \in \mathbf{E}^*$, $f_{b,c}^w(a)$ and $f_{b,c}^w(1)$ have the same distribution as $(b, c)$ runs through all elements in $\mathbf{E}^* \times \mathbf{F}^*$. Thus, the distribution of $f_{b,c}^w(1)$ can be obtained as Eq. (57).

(2) The case for even $n/2$ can be similarly proved. ∎

For odd $n/2$, by analyzing the rank of $f_{b,c}(x)$ and applying Proposition 17, the following results can be obtained. Another proof is provided here.

*Proposition 18:* Assume $n/2$ is odd. When $(b,c)$ runs through all elements in $\mathbf{E} \times \mathbf{F}$, the joint distribution of $f_{b,c}^w(1)$ and $f_{1,c}^w(0)$ is given by

$$
\begin{cases}
\pm 2^{n/2+1}, & (2^{n-3} \pm 2^{n/2-2})(2^{n/2+1}-1)/3 \text{ times}; \\
0, & 2^{3n/2-1} - 2^{n-2} + 1 \text{ times}; \\
2^{n/2}, & (2^{3n/2-1} + 2^n + 2^{n/2+1})/3 \text{ times}; \\
-2^{n/2}, & (2^{3n/2-1} + 2^{n/2} - 3)/3 \text{ times}.
\end{cases}
\tag{61}
$$

*Proof:* Since $\gcd(2^k+1, 2^n-1) = 1$, for any $b \in \mathbf{E}^*$, there is $e \in \mathbf{E}^*$ such that $b = e^{2^k+1}$. It is easy to check

$$f_{b,c}^w(1) = f_{1,\,ce^{-2^{n/2}-1}}^w(e^{-1}).$$

Let $b_i = e_i^{2^k+1}$ for $i = 1, 2$, one has

$$(c_1 e_1^{-2^{n/2}-1}, e_1^{-1}) = (c_2 e_2^{-2^{n/2}-1}, e_2^{-1})$$

if and only if $(b_1, c_1) = (b_2, c_2)$. Thus, the joint distribution of $f_{b,c}^w(1)$ and $f_{1,c}^w(0)$ as $(b,c)$ runs through all elements in $\mathbf{E}^* \times \mathbf{F}$, is the same as the Fourier transform value distribution of $f_{1,c}(x)$ as $c$ runs through all elements in $\mathbf{F}$.

By Corollary 12, there are exactly $(2^{n+1} - 2^{n/2+1} - 4)/3$ elements $b \in \mathbf{E}^*$ such that the rank of $f_{b,1}(x)$ is $n-2$. Let $y = ex$, then $f_{b,1}(x) = f_{1,e^{-2^{n/2}-1}}(y)$. Thus, the rank of $f_{b,1}(x)$ is the same as that of $f_{1,e^{-2^{n/2}-1}}(x)$. When $b$ runs through all elements in $\mathbf{E}^*$, $e^{-2^{n/2}-1}$ runs through all elements in $\mathbf{F}^*$ for $2^{n/2}+1$ times. Hence, there are exactly

$$(2^{n+1} - 2^{n/2+1} - 4)/(3(2^{n/2}+1)) = 2(2^{n/2}-2)/3$$

elements $c \in \mathbf{F}^*$ such that the rank of $f_{1,c}(x)$ is $n-2$.

Therefore, the distribution of the Fourier transform values of $f_{1,c}(x)$ as $c$ runs through all elements in $\mathbf{F}^*$ is

$$
\begin{cases}
\pm 2^{n/2+1}, & (2^{n-3} \pm 2^{n/2-2})(2^{n/2+1}-4)/3 \text{ times} \\
0, & 2^{n-1}(2^{n/2}-2) \text{ times} \\
\pm 2^{n/2}, & (2^{n-1} \pm 2^{n/2-1})(2^{n/2}+1)/3 \text{ times}.
\end{cases}
$$

 

By Proposition 5 and Theorem 1, the distribution of the Fourier transform values of $f_{1,0}(x)$ is given by

$$
\begin{cases}
\pm 2^{n/2+1}, & 2^{n-3} \pm 2^{n/2-2} \text{ times} \\
0, & 2^n - 2^{n-2} \text{ times.}
\end{cases}
$$

Combining the above two distributions, $f_{0,0}^w(1) = 0$, and the distribution of $f_{0,c}^w(1)$ for $c \in \mathbf{F}^*$ given in Proposition 6, we obtain the distribution in Eq. (61). ∎

Consider the case for even $n/2$.

*Lemma 19:* Assume $n/2$ is even. Let $\Gamma$ and $\Delta$ be defined as in Eq. (26). For $i = 0, 1, \cdots, (2^n - 4)/3$ and $j = 1, 2$, as $(b, c)$ and $(b', c')$ run through all elements in $\alpha^{3i}\Gamma \times \Delta$ and $\alpha^{3i}\Gamma \times \beta^{(3-j)(2^{n/2}-1)/3}\Delta$, respectively, $f_{b,c}^w(0)$ and $f_{b',c'}^w(0)$ have the same distribution. In particular, one has

(1) When $(b, c)$ runs through all elements in $\mathbf{E}^* \times \Delta$ for exactly 3 times, the distribution of $f_{b,c}^w(0)$ is the same as that of $f_{b',c'}^w(0)$ as $(b', c')$ runs through all elements in $\mathbf{E}^* \times \mathbf{F}^*$;

(2) When $(b, c)$ runs through all elements in $\Gamma \times \Delta$ for exactly 3 times, the distribution of $f_{b,c}^w(0)$ is the same as that of $f_{b',c'}^w(0)$ as $(b', c')$ runs through all elements in $\Gamma \times \mathbf{F}^*$.

*Proof:* Since

$$
f_{b,c}(\alpha^l x) = f_{b\alpha^{l(2^k+1)}, c\beta^l}(x),
$$

the rank of $f_{b,c}(x)$ is the same as that of $f_{b\alpha^{l(2^k+1)}, c\beta^l}(x)$. Furthermore, for $(b, c) \in \alpha^{3i}\Gamma \times \Delta$ and $(b', c') \in \alpha^{3i}\Gamma \times \beta^{(3-j)(2^{n/2}-1)/3}\Delta$, one has

$$
f_{b\alpha^{l(2^k+1)}, c\alpha^{l(2^{n/2}+1)}}(x) = f_{b',c'}(x)
$$

if and only if

$$
l = j(2^n - 1)/3, \ \ b' = b, \ \ c' = c\beta^l.
$$

Since $n/2$ is even, one has $2^{n/2} + 1 \equiv 2 \bmod 3$ and then

$$
l = j(2^{n/2} - 1)(2^{n/2} + 1)/3 \equiv 2j(2^{n/2} - 1)/3 \equiv (3 - j)(2^{n/2} - 1)/3 \bmod (2^{n/2} - 1).
$$

Thus, $c' \in \beta^{(3-j)(2^{n/2}-1)/3}\Delta$ for any $c \in \Delta$. ∎

By Lemma 19, when $(b, c)$ and $(b', c')$ take over $\Gamma \times \Delta$ and $\Gamma \times \mathbf{F}^*$, respectively, the distribution of $f_{b,c}^w(0)$ is determined by calculating the distribution of $f_{b',c'}^w(0)$. The latter can be obtained by the following lemma.

*Lemma 20:* Assume $n/2$ is even. The distribution of $f_{b,c}^w(0)$ as $(b,c)$ runs through all elements in $\Gamma \times \mathbf{F}^*$ for exactly $(2^n-1)/3$ times, is the same as that of $f_{b',c'}^w(0)$ as $(b',c')$ runs through all elements in $\mathbf{E}^* \times \mathbf{F}^*$.

*Proof:* Since $\gcd(2^k+1, 2^n-1) = 3$, there is an element $e \in \mathbf{E}$ such that $e^{2^k+1} = \alpha^3$. One has

$$f_{b,c}(e^i x) = f_{b\alpha^{3i}, ce^{i(2^{n/2}+1)}}(x),$$

and so,

$$f_{b,c}^w(0) = f_{b\alpha^{3i}, ce^{i(2^{n/2}+1)}}^w(0).$$

On the other hand, as $(b,c)$ runs through all elements in $\Gamma \times \mathbf{F}^*$, $(b\alpha^{3i}, ce^{i(2^{n/2}+1)})$ runs through all elements in $\alpha^{3i}\Gamma \times \mathbf{F}^*$. This proves the lemma. ∎

The following proposition derived from Lemmas 19 and 20 is an analogy of Proposition 18.

*Proposition 21:* Assume $n/2$ is even. When $(b,c)$ runs through all elements in $\mathbf{E} \times \mathbf{F}$ for exactly $(2^n + 2^{n/2} - 1)$ times and $(b',c')$ runs through all elements in

$$\bigcup_{(\zeta_1, \eta_1) \in \Gamma \times \Delta} (\{\zeta_1\} \times \mathbf{F}\backslash\{\eta_1\} \bigcup \mathbf{E} \times \{\eta_1\})$$

for exactly once, the joint distribution of $f_{b,c}^w(1)$ and $f_{b',c'}^w(0)$ is given by

$$\begin{cases} 2^{n/2+1}, & (2^n + 2^{n/2} - 1)(2^{n/2+1} - 1)(2^{n-3} + 2^{n/2-2})/3 \text{ times;} \\ -2^{n/2+1}, & (2^{5n/2-2} - 3 \cdot 2^{2n-3} - 5 \cdot 2^{3n/2-3} - 2^{n-3} - 5 \cdot 2^{n/2-2} + 4)/3 \text{ times;} \\ 0, & 2^{5n/2-1} + 2^{2n-2} - 3 \cdot 2^{3n/2-2} + 5 \cdot 2^{n-2} - 1 \text{ times;} \\ 2^{n/2}, & (2^{5n/2-1} + 3 \cdot 2^{2n-1} + 5 \cdot 2^{3n/2-1} - 2^n - 2^{n/2+2} + 2)/3 \text{ times;} \\ -2^{n/2}, & (2^{5n/2-1} + 2^{2n-1} + 2^{3n/2-1} - 2^{n+1} - 2^{n/2})/3 \text{ times.} \end{cases} \tag{62}$$

*Proof:* By Propositions 5, 6 and 17(2), the distribution of $f_{b,c}^w(1)$ as $(b,c)$ runs through all elements in $\mathbf{E} \times \mathbf{F}$, is given by

$$f_{b,c}^w(1) = \begin{cases} 2^{n/2+1}, & (2^{n/2+1} - 1)(2^{n-3} + 2^{n/2-2})/3 \text{ times;} \\ -2^{n/2+1}, & (2^{3n/2-2} - 5 \cdot 2^{n-3} - 2^{n/2-2} + 1)/3 \text{ times;} \\ 0, & 2^{3n/2-1} - 2^{n-2} - 2^{n/2-1} + 1 \text{ times;} \\ 2^{n/2}, & (2^{3n/2-1} + 2^n + 2^{n/2} - 1)/3 \text{ times;} \\ -2^{n/2}, & (2^{3n/2-1} + 2^{n/2} - 3)/3 \text{ times.} \end{cases}$$

For any $(\zeta_1, \eta_1) \in \Gamma \times \Delta$, the set $\{\zeta_1\} \times \mathbf{F}$ consists of two parts: $\{\zeta_1\} \times \mathbf{F} \backslash \{\eta_1\}$ and $\{\zeta_1\} \times \{\eta_1\}$. Therefore, Combining Lemma 19, 20, and Propositions 5, 6, and 17(2), one has as $(b', c')$ runs through all elements in

$$\bigcup_{(\zeta_1, \eta_1) \in \Gamma \times \Delta} (\{\zeta_1\} \times \mathbf{F} \backslash \{\eta_1\} \bigcup \mathbf{E} \times \{\eta_1\}),$$

the distribution of $f_{b', c'}^w(0)$ is given by

$$f_{b', c'}^w(0) = \begin{cases} 2^{n/2+1}, & 0 \text{ times;} \\ -2^{n/2+1}, & (2^{3n/2-1} - 3 \cdot 2^{n-1} - 5 \cdot 2^{n/2-1} + 5)/3 \text{ times;} \\ 0, & (2^n + 2^{n/2} - 3)2^{n/2-1} \text{ times;} \\ 2^{n/2}, & (2^n + 2^{n/2} - 1)(2^{n/2} - 1)/3 \text{ times;} \\ -2^{n/2}, & 2^{n/2} - 1 \text{ times.} \end{cases}$$

Combining the above two distributions gives Eq. (62). ∎

## V. New families of binary sequences

In this section, for even $n$, we generalize the construction for the large set of Kasami sequences. New families of binary sequences with six-valued correlation and family sizes $2^{3n/2} + 2^{n/2}$ or $2^{3n/2} + 2^{n/2} - 1$ are obtained. Their correlation distributions are completely determined by applying the results in previous sections.

*Definition 22:* Let $k$ be an integer satisfying the condition in Eq. (35). The family $\mathcal{F}^k$ of binary sequences of period $2^n - 1$ is defined by

$$\mathcal{F}^k = \{\{s_{\gamma\delta}(t)\}_{t=0}^{2^n-2} \,|\, \gamma \in \mathbf{E}, \delta \in \mathbf{F}\} \bigcup \{\{s_{\zeta\eta}(t)\}_{t=0}^{2^n-2} \,|\, \zeta \in \Gamma, \ \eta \in \Delta\},$$

where

$$s_{\gamma\delta}(t) = tr_1^n(\alpha^t + \gamma\alpha^{t(2^k+1)}) + tr_1^{n/2}(\delta\alpha^{t(2^{n/2}+1)}),$$

$$s_{\zeta\eta}(t) = tr_1^n(\zeta\alpha^{t(2^k+1)}) + tr_1^{n/2}(\eta\alpha^{t(2^{n/2}+1)}),$$

and $\Gamma$ and $\Delta$ are defined in Eq. (25) and Eq. (26).

For convenience, set

$$\mathcal{F}_1^k = \{\{s_{\gamma\delta}(t)\}_{t=0}^{2^n-2} \,|\, \gamma \in \mathbf{E}, \delta \in \mathbf{F}\}, \quad \mathcal{F}_2^k = \{\{s_{\zeta\eta}(t)\}_{t=0}^{2^n-2} \,|\, \zeta \in \Gamma, \ \eta \in \Delta\}.$$

The correlation function between $\{s_{\gamma_1\delta_1}\}$ and $\{s_{\gamma_2\delta_2}\}$ is

$$R_{\gamma_1\delta_1,\gamma_2\delta_2}(\tau) = \sum_{x\in\mathbf{E}}(-1)^{g(x)} - 1,$$

where

$$\begin{aligned}
g(x) &= tr_1^n(x+\gamma_1 x^{2^k+1}) + tr_1^{n/2}(\delta_1 x^{2^{n/2}+1}) \\
&\quad + tr_1^n(\alpha^\tau x + \gamma_2\alpha^{\tau(2^k+1)}x^{2^k+1}) + tr_1^{n/2}(\delta_2\alpha^{\tau(2^{n/2}+1)}x^{2^{n/2}+1}) \ , \\
&= tr_1^n(a_1 x + b_1 x^{2^k+1}) + tr_1^{n/2}(c_1 x^{2^{n/2}+1})
\end{aligned}$$

where $a_1 = 1+\alpha^\tau$, $b_1 = \gamma_1+\gamma_2\alpha^{\tau(2^k+1)}$ and $c_1 = \delta_1+\delta_2\alpha^{\tau(2^{n/2}+1)}$. Thus,

$$R_{\gamma_1\delta_1,\gamma_2\delta_2}(\tau) = f_{b_1,c_1}^w(a_1) - 1 \tag{63}$$

Similarly, the correlation functions $R_{\gamma\delta,\zeta\eta}(\tau)$, $R_{\zeta\eta,\gamma\delta}(\tau)$ and $R_{\zeta_1\eta_1,\zeta_2\eta_2}(\tau)$ are given by

$$\begin{cases}
R_{\gamma\delta,\zeta\eta}(\tau) = f_{b_2,c_2}^w(a_2) - 1 \\
R_{\zeta\eta,\gamma\delta}(\tau) = f_{b_3,c_3}^w(a_3) - 1 \\
R_{\zeta_1\eta_1,\zeta_2\eta_2}(\tau) = f_{b_4,c_4}^w(a_4) - 1
\end{cases} \tag{64}$$

where

$$\begin{aligned}
a_2 &= 1, \ b_2 = \gamma+\zeta\alpha^{\tau(2^k+1)}, \ c_2 = \delta+\eta\alpha^{\tau(2^{n/2}+1)}; \\
a_3 &= \alpha^\tau, \ b_3 = \zeta+\gamma\alpha^{\tau(2^k+1)}, \ c_3 = \eta+\delta\alpha^{\tau(2^{n/2}+1)}; \\
a_4 &= 0, \ b_4 = \zeta_1+\zeta_2\alpha^{\tau(2^k+1)}, \ c_4 = \eta_1+\eta_2\alpha^{\tau(2^{n/2}+1)}.
\end{aligned}$$

By Eq. (63) and Eq. (64) together with the Fourier transform values of $f_{b,c}(a)$ in Section IV, the correlation distribution of $\mathcal{F}^k$ can be obtained as follows.

*Theorem 23:* Assume $n/2$ is odd. The distribution of the correlation values of the family $\mathcal{F}^k$ is given by

$$\begin{cases}
2^n-1, & 2^{3n/2}+2^{n/2} \text{ times;} \\
-1, & 2^{n/2+1}(2^{7n/2-2}-2^{3n-3}+2^{2n-1}-2^{3n/2-1}+2^{n-2}-1) \text{ times;} \\
2^{n/2}-1, & (2^{4n-1}+2^{7n/2}+2^{3n+1}-2^{2n}-2^{3n/2+1}-2^{n+2})/3 \text{ times;} \\
-2^{n/2}-1, & (2^{4n-1}+2^{3n}-3\cdot2^{5n/2}+2^{2n+1}-3\cdot2^{3n/2}+2^n+3\cdot2^{n/2})/3 \text{ times;} \\
\pm2^{n/2+1}-1, & 2^{n/2+1}(2^{n-3}\pm2^{n/2-2})(2^{2n-1}-1)(2^{n/2+1}-1)/3 \text{ times.}
\end{cases} \tag{65}$$

*Proof:* By Eq. (63), for given $\tau$ and $i=1,2$, as $(\gamma_i,\delta_i)$ runs through all elements in $\mathbf{E}\times\mathbf{F}$ for exactly once, $(b_1,c_1)$ runs through all elements in $\mathbf{E}\times\mathbf{F}$ for exactly $2^{3n/2}$ times.

Similarly, by Eq. (64), for each fixed $\tau$, $(b_2, c_2)$ runs through all elements in $\mathbf{E} \times \mathbf{F}$ for exactly $2^{n/2}$ times as $(\gamma, \zeta)$ and $(\delta, \eta)$ run through all elements in $\mathbf{E} \times \Gamma$ and $\mathbf{F} \times \Delta$, respectively. And as $\tau$ varies from 0 to $2^n - 2$, $a_1$ runs through all elements in $\mathbf{E} \backslash \{1\}$ once.

Thus, when $(b, c)$ runs through all elements in $\mathbf{E} \times \mathbf{F}$ for exactly $2^{n/2}$ times, by Proposition 13, the distribution of $f_{b,c}^w(1) - 1$ and the correlation values of sequences between $\mathcal{F}_1^k$ and $\mathcal{F}_1^k \cup \mathcal{F}_2^k$ is given by

$$\begin{cases} \pm 2^{n/2+1} - 1, & 2^{3n/2}(2^{n/2} - 1)(2^{n-3} \pm 2^{n/2-2})(2^{n+1} + 2^{n/2} - 1)/3 \text{ times}; \\ -1, & 2^{3n/2}(2^{n/2} - 1)(2^{2n-1} + 2^{3n/2-2} - 2^{n-2} + 2^{n/2} + 1) \text{ times}; \\ \pm 2^{n/2} - 1, & 2^{3n/2}(2^{n/2} - 1)(2^{n-1} \pm 2^{n/2-1})(2^n + 2^{n/2+1} + 4)/3 \text{ times}; \\ 2^n - 1, & 2^{3n/2} \text{ times}. \end{cases}$$

Given a $\tau$, by Eq. (64), $(b_3, c_3)$ runs through all elements in $\mathbf{E} \times \mathbf{F}$ for exactly $2^{n/2}$ times as $(\gamma, \zeta)$ and $(\eta, \delta)$ run through all elements in $\mathbf{E} \times \Gamma$ and $\Delta \times \mathbf{F}$, respectively.

By assumption, $n/2$ is odd, and $\zeta_1 = \zeta_2 = 1$. For fixed $\eta_1$, by Eq. (64), $(b_4, c_4)$ runs through all elements in $\mathbf{E} \backslash \{1\} \times \mathbf{F}$ for exactly once as $\eta_2$ runs through all elements in $\mathbf{F}$ and $\tau$ varies from 0 to $2^n - 2$. Meanwhile, as $\tau$ varies from 0 to $2^n - 2$, $a_3$ runs through all elements in $\mathbf{E}^*$ for exactly once.

Thus, when $c$ runs through all elements in $\mathbf{F}$ for exactly $2^{n/2}$ times, by Proposition 13, the distribution of $f_{1,c}^w(0) - 1$ and the correlation values of sequences between $\mathcal{F}_2^k$ and $\mathcal{F}_1^k \cup \mathcal{F}_2^k$ is given by

$$\begin{cases} \pm 2^{n/2+1} - 1, & 2^{n/2}(2^{n/2} - 1)(2^{n-3} \pm 2^{n/2-2})(2^{n+1} + 2^{n/2} - 1)/3 \text{ times}; \\ -1, & 2^{n/2}(2^{n/2} - 1)(2^{2n-1} + 2^{3n/2-2} - 2^{n-2} + 2^{n/2} + 1) \text{ times}; \\ \pm 2^{n/2} - 1, & 2^{n/2}(2^{n/2} - 1)(2^{n-1} \pm 2^{n/2-1})(2^n + 2^{n/2+1} + 4)/3 \text{ times}; \\ 2^n - 1, & 2^{n/2} \text{ times}. \end{cases}$$

Combining the above two distributions together with Proposition 18, the distribution of correlation values of $\mathcal{F}^k$ is obtained as Eq. (65). ∎

The imbalance $I(s)$ of a binary sequence $s$ is the difference of the times zeros and ones appear in $s$. The imbalance of sequences in $\mathcal{F}^k$ is distributed as follows.

*Proposition 24:* Assume $n/2$ is odd. The imbalance distribution of sequences in $\mathcal{F}^k$ is

$$I(s) = \begin{cases} \pm 2^{n/2+1} - 1, & (2^{n-3} \pm 2^{n/2-2})(2^{n/2+1} - 1)/3 \text{ times}; \\ -1, & 2^{3n/2-1} - 2^{n-2} + 1 \text{ times}; \\ 2^{n/2} - 1, & (2^{3n/2-1} + 2^n + 2^{n/2+1})/3 \text{ times}; \\ -2^{n/2} - 1, & (2^{3n/2-1} + 2^{n/2} - 3)/3 \text{ times}. \end{cases} \tag{66}$$

*Proof:* It is sufficient to determine the joint distribution of $f_{b,c}^w(1)$ and $f_{1,c}^w(0)$ as $(b,c)$ runs through all elements in $\mathbf{E} \times \mathbf{F}$. By Proposition 18, the imbalance distribution is easy to obtain as Eq. (66). ∎

*Theorem 25:* Assume $n/2$ is even. The distribution of the correlation values of the family $\mathcal{F}^k$ is given as

$$\begin{cases} 2^n - 1, & 2^{3n/2} + 2^{n/2} - 1 \text{ times}; \\ -1, & 2^{4n-1} - 2^{7n/2-2} - 2^{2n-1} + 3 \cdot 2^{3n/2-1} - 5 \cdot 2^{n-1} - 2^{n/2} + 2 \text{ times}; \\ 2^{n/2} - 1, & (2^{4n-1} + 2^{7n/2} + 2^{3n+1} - 2^{5n/2} - 3 \cdot 2^{2n} - 5 \cdot 2^{3n/2} + 3 \cdot 2^{n/2+1} - 2)/3 \text{ times}; \\ -2^{n/2} - 1, & (2^{4n-1} + 2^{3n} - 2^{5n/2+2} + 2^{2n+1} - 2^{3n/2+2} + 7 \cdot 2^n - 2^{n/2})/3 \text{ times}; \\ 2^{n/2+1} - 1, & (2^{4n-2} + 3 \cdot 2^{7n/2-3} - 2^{3n-2} - 2^{5n/2-1} - 5 \cdot 2^{2n-2} + 2^{3n/2-2} + \\ & \qquad\qquad 5 \cdot 2^{n-2} - 2^{n/2-1})/3 \text{ times}; \\ -2^{n/2+1} - 1, & (2^{4n-2} - 5 \cdot 2^{7n/2-3} + 2^{3n-2} - 2^{5n/2-1} + 3 \cdot 2^{2n-2} + 5 \cdot 2^{3n/2-2} - \\ & \qquad\qquad 3 \cdot 2^{n-2} + 3 \cdot 2^{n/2-1} - 4)/3 \text{ times}. \end{cases} \tag{67}$$

*Proof:* Similar to the case for odd $n/2$, when $(b,c)$ runs through all elements in $\mathbf{E} \times \mathbf{F}$ for $(2^n + 2^{n/2} - 1)$ times, the distribution of $f_{b,c}^w(1) - 1$ and correlation values of sequences between $\mathcal{F}_1^k$ and $\mathcal{F}_1^k \cup \mathcal{F}_2^k$ is given by

$$\begin{cases} \pm 2^{n/2+1} - 1, & 2^{3n/2}(2^{n/2} - 1)(2^{n-3} \pm 2^{n/2-2})(2^{n+1} + 2^{n/2} - 1)/3 \text{ times} \\ -1, & 2^{3n/2}(2^{n/2} - 1)(2^{2n-1} + 2^{3n/2-2} - 2^{n-2} + 2^{n/2} + 1) \text{ times} \\ \pm 2^{n/2} - 1, & 2^{3n/2}(2^{n/2} - 1)(2^{n-1} \pm 2^{n/2-1})(2^n + 2^{n/2+1} + 4)/3 \text{ times} \\ 2^n - 1, & 2^{3n/2} \text{ times}. \end{cases}$$

The set $\mathbf{E} \times \mathbf{F}$ can be divided into three disjoint parts: $\{\zeta_1\} \times \mathbf{F} \backslash \{\eta_1\}$, $\mathbf{E} \backslash \{\zeta_1\} \times \mathbf{F} \backslash \{\eta_1\}$, and $\mathbf{E} \times \{\eta_1\}$. When $(b,c)$ runs through all elements in

$$\bigcup_{(\zeta_1, \eta_1) \in \Gamma \times \Delta} (\{\zeta_1\} \times \mathbf{F} \backslash \{\eta_1\} \bigcup \mathbf{E} \times \{\eta_1\})$$

for exactly once, the distribution of $f^w_{b,c}(0) - 1$ and correlation values of sequences between $\mathcal{F}^k_2$ and $\mathcal{F}^k_1 \cup \mathcal{F}^k_2$ is given by

$$
\begin{cases}
\pm 2^{n/2+1} - 1, & (2^{n/2} - 1)(2^{n/2} - 1)(2^{n-3} \pm 2^{n/2-2})(2^{n+1} + 2^{n/2} - 1)/3 \text{ times}; \\
-1, & (2^{n/2} - 1)(2^{n/2} - 1)(2^{2n-1} + 2^{3n/2-2} - 2^{n-2} + 2^{n/2} + 1) \text{ times}; \\
\pm 2^{n/2} - 1, & (2^{n/2} - 1)(2^{n/2} - 1)(2^{n-1} \pm 2^{n/2-1})(2^n + 2^{n/2+1} + 4)/3 \text{ times}; \\
2^n - 1, & 2^{n/2} - 1 \text{ times}.
\end{cases}
$$

By Proposition 21, the distribution of correlation values of family $\mathcal{F}^k$ is obtained as Eq. (67). ∎

*Proposition 26:* Assume $n/2$ is even. The imbalance distribution of sequences in $\mathcal{F}^k$ is

$$
I(s) = \begin{cases}
2^{n/2+1} - 1, & (2^{n/2+1} - 1)(2^{n-3} + 2^{n/2-2})/3 \text{ times}; \\
-2^{n/2+1} - 1, & (2^{3n/2-2} - 5 \cdot 2^{n-3} + 2^{n/2-2} - 1)/3 \text{ times}; \\
-1, & 2^{3n/2-1} - 2^{n-2} + 1 \text{ times}; \\
2^{n/2} - 1, & (2^{3n/2-1} + 2^n + 2^{n/2+1} - 2)/3 \text{ times}; \\
-2^{n/2} - 1, & (2^{3n/2-1} + 2^{n/2} - 3)/3 \text{ times}.
\end{cases}
\tag{68}
$$

*Proof:* It is sufficient to determine the distribution of $f^w_{b,c}(1)$ and $f^w_{b',c'}(0)$, as $(b,c)$ and $(b',c')$ run through all elements in $\mathbf{E} \times \mathbf{F}$ and $\Gamma \times \Delta$, respectively. By Propositions 5, 6, 17 and Lemmas 19 and 20, the imbalance distribution is obtained as Eq. (68).

Table I summarizes the properties of some families with low correlation.

Below is an example obtained through computer search.

*Example 27:* (1) Let $n = 6$ and $k = 2$. The correlation distribution of $\mathcal{F}^2$ is given by

$$63, \quad 520 \text{ times}; \quad -1, \quad 7893232 \text{ times}; \quad 7, \quad 3668224 \text{ times};$$

$$-9, \quad 2853064 \text{ times}; \quad 15, \quad 1637600 \text{ times}; \quad -17, \quad 982560 \text{ times}.$$

(2) Let $n = 4$ and $k = 1$. The correlation distribution of $\mathcal{F}^1$ is given by

$$15, \quad 67 \text{ times}; \quad -1, \quad 28598 \text{ times}; \quad 3, \quad 18418 \text{ times};$$

$$-5, \quad 11044 \text{ times}; \quad 7, \quad 6902 \text{ times}; \quad -9, \quad 2306 \text{ times}.$$

These distributions agree with the corresponding results given by Eq. (65) and Eq. (67).

Take

$$
k = \begin{cases}
m + 1, & \text{for } m \text{ odd}; \\
m, & \text{for } m \text{ even}.
\end{cases}
\tag{69}
$$

TABLE I

SOME FAMILIES OF BINARY SEQUENCES OF PERIOD $2^n - 1$ WITH LOW CORRELATION.

| Family | $n$ | Family size | $R_{\max}$ |
|---|---|---|---|
| Gold and Gold-like sequences | $n = 2m + 1$ | $2^n + 1$ | $1 + 2^{(n+1)/2}$ |
| GKW and GKW-like* sequences | $n = me, m$ odd | $2^n + 1$ | $1 + 2^{(n+e)/2}$ |
| Large set of Kasami sequences | $n = 4m + 2$ | $2^{3n/2} + 2^{n/2}$ | $1 + 2^{n/2+1}$ |
| | $n = 4m$ | $2^{3n/2} + 2^{n/2} - 1$ | |
| The proposed families $\mathcal{F}^k$ | $n = 4m + 2$, $\gcd(n, k) = 2$ | $2^{3n/2} + 2^{n/2}$ | $1 + 2^{n/2+1}$ |
| | $n = 4m$, $\gcd(n, k) = 1$ | $2^{3n/2} + 2^{n/2} - 1$ | |

* GKW-like sequences are those sequences constructed in [9].

for $n = 4m + 2 \geq 10$, and take $k = 1$ for $n = 4m \geq 8$. Then, $k \neq n/2 \pm 1$. Thus, for any $n \geq 8$, there is parameter $k$ such that $\mathcal{F}^k$ is different from the large set of Kasami sequences. The proposed families are generalizations of the large set of Kasami sequences.

## VI. CONCLUSIONS

New families of binary sequences of period $2^n - 1$ have been proposed in this paper for even $n$. They are generalizations of the large set of Kasami sequences. These families achieve low correlation. Through studying a class of equations over finite fields and determining the distribution of Fourier transforms for the functions defined in Eq. (13), the correlation distributions of the proposed families are completely determined.